\documentclass{pasj01}
\usepackage{url}
\Received{$\langle$reception date$\rangle$}
\Accepted{$\langle$acception date$\rangle$}
\Published{$\langle$publication date$\rangle$}

\RequirePackage{color}

\begin{document}

\title{Bo\"otes IV: A New Milky Way Satellite Discovered in
the Subaru Hyper Suprime-Cam Survey and Implications for the Missing Satellite Problem}

\author{Daisuke~Homma\altaffilmark{1}, Masashi~Chiba\altaffilmark{2}, 
Yutaka~Komiyama\altaffilmark{1,3}, Masayuki~Tanaka\altaffilmark{1}, 
Sakurako~Okamoto\altaffilmark{1,7}, Mikito~Tanaka\altaffilmark{4},
Miho~N.~Ishigaki\altaffilmark{5,2}, Kohei~Hayashi\altaffilmark{6},
Nobuo~Arimoto\altaffilmark{7,10},
Scott~G.~Carlsten\altaffilmark{8},
Robert~H.~Lupton\altaffilmark{8}, Michael~A.~Strauss\altaffilmark{8}, 
Satoshi~Miyazaki\altaffilmark{1,3}, 
Gabriel~Torrealba\altaffilmark{9}, Shiang-Yu~Wang\altaffilmark{9}, and
Hitoshi~Murayama\altaffilmark{5}
}

\altaffiltext{1}{National Astronomical Observatory of Japan, 2-21-1 Osawa, Mitaka, Tokyo 181-8588, Japan}
\altaffiltext{2}{Astronomical Institute, Tohoku University, Aoba-ku, Sendai 980-8578, Japan}
\altaffiltext{3}{The Graduate University for Advanced Studies, Osawa 2-21-1, Mitaka, Tokyo 181-8588, Japan}
\altaffiltext{4}{Department of Advanced Sciences, Faculty of Science and Engineering, Hosei University, 184-8584 Tokyo, Japan}
\altaffiltext{5}{Kavli Institute for the Physics and Mathematics of the Universe (WPI),
  The University of Tokyo, Kashiwa, Chiba 277-8583, Japan}
\altaffiltext{6}{ICRR, The University of Tokyo, Kashiwa, Chiba 277-8583, Japan}
\altaffiltext{7}{Subaru Telescope, National Astronomical Observatory of Japan, 650 North A'ohoku Place,
Hilo, HI 96720, USA}
\altaffiltext{8}{Princeton University Observatory, Peyton Hall, Princeton, NJ 08544, USA}
\altaffiltext{9}{Institute of Astronomy and Astrophysics, Academia Sinica, Taipei, 10617, Taiwan}
\altaffiltext{10}{Astronomy Program, Department of Physics and Astronomy,
Seoul National University, 599 Gwanak-ro, Gwanak-gu, Seoul, 151-742, Korea}
\email{chiba@astr.tohoku.ac.jp}

\KeyWords{galaxies: dwarf --- galaxies: individual (Bo\"otes~IV) --- globular clusters: individual (HSC~1)
--- Local Group}

\maketitle

\begin{abstract}
We report on the discovery of a new Milky Way (MW) satellite in Bo\"otes based on data from the on-going
Hyper Suprime-Cam (HSC) Subaru Strategic Program (SSP). This satellite, named Bo\"otes~IV, is the third
ultra-faint dwarf that we have discovered in the HSC-SSP. We have identified a statistically
significant (32.3$\sigma$) overdensity of stars having characteristics of
a metal-poor, old stellar population. The distance to this stellar system is
$D_{\odot}=209^{+20}_{-18}$~kpc with a $V$-band absolute magnitude of
$M_V=-4.53^{+0.23}_{-0.21}$~mag. Bo\"otes~IV has a half-light radius of $r_h=462^{+98}_{-84}$~pc
and an ellipticity of $0.64^{+0.05}_{-0.05}$, which clearly suggests 
that this is a dwarf satellite galaxy.
We also found another overdensity that appears to be a faint globular
cluster with $M_V=-0.20^{+0.59}_{-0.83}$ mag and $r_h=5.9^{+1.5}_{-1.3}$~pc
located at $D_{\odot}=46^{+4}_{-4}$~kpc.
Adopting the recent prediction for the total population of satellites in a MW-sized halo
by \citet{Newton2018}, which combined the characteristics of the observed satellites 
by the Sloan Digital Sky Survey and the Dark Energy Survey with the subhalos obtained in $\Lambda$CDM models,
we estimate that there should be about two MW satellites at $M_V\le0$ in the $\sim676$~deg$^2$ covered
by HSC-SSP, whereas that area includes six satellites (Sextans, Leo~IV, Pegasus~III, Cetus~III,
Virgo~I and Bo\"otes~IV). Thus, the observed number of satellites
is larger than the theoretical prediction. On the face of it, we have a problem of too many satellites,
instead of the well-known missing satellites problem whereby the $\Lambda$CDM theory overpredicts
the number of satellites in a MW-sized halo.
This may imply that the models need more refinements for the assignment of subhalos to
satellites such as considering those found by the current deeper survey.
Statistically more robust constraints on this issue will be brought by further surveys
of HSC-SSP over the planned $\sim1,400$~deg$^2$ area.
\end{abstract}


\section{Introduction}

The $\Lambda$-dominated cold dark matter ($\Lambda$CDM) model has become the standard
cosmological model.  A strong, immediate prediction of the model is that objects in
the Universe grow hierarchically; they become progressively more massive with time through
accretion and mergers.
The $\Lambda$CDM model reproduces the observed cosmic large-scale structure
on scales  $\gtrsim 1$~Mpc extremely well  (e.g., \cite{Tegmark2004}).  However,
the model is not perfect, and there is extensive discussion in the literature of several
persistent problems on smaller scales.  These problems include the core or cusp problem (e.g.,
\cite{Moore1994,Burkert1995,de Blok2001,Gilmore2007,Oh2011}),
the so-called too-big-to-fail problem (e.g., \cite{Boylan-Kolchin2011,Boylan-Kolchin2012},
the satellite alignment problem (e.g.,
\cite{Kroupa2005,McConnachie2006,Ibata2013,Pawlowski2012,Pawlowski2013,Pawlowski2015}),
and the missing satellites problem \citep{Klypin1999,Moore1999}.
The missing satellites problem, namely that a MW-sized halo is predicted to have significantly
more subhalos than the observed number of dwarf satellites,
simply refers to the satellite abundance, whereas the others are related
to the internal density profiles of satellite galaxies and their dynamics in a MW-sized halo.
It is thus important to fully understand the missing satellite problem in order to
address the higher order issues.

A few possible solutions to the missing satellites problem have been proposed in
the literature.  One is to invoke baryon physics to suppress star formation in
low-mass subhalos, thus making the majority of these low-mass halos starless
\citep{Sawala2016a,Sawala2016b}.
Another is to suggest that our catalogs of dwarf satellites is incomplete; we are not
counting all the satellites due to
observational constraints such as limited survey area and depth.  These constraints may
result in a significantly underestimated satellite abundance \citep{Koposov2008,Tollerud2008}.
However, the observational situation has dramatically improved in recent years thanks
largely to modern massive imaging surveys.
The Sloan Digital Sky Survey (SDSS; \cite{York2000}), the Dark Energy Survey (DES; \cite{Abbott2016}),
and the Pan-STARRS~1 (PS1) 3$\pi$ survey \citep{Chambers2016} have covered wide areas of sky
and discovered a large number of so-called ultra-faint dwarf galaxies (UFDs), whose
$V$-band absolute magnitudes ($M_V$) are fainter than about $-8$~mag
(e.g., \cite{Willman2005,Sakamoto2006,Belokurov2006,
Laevens2014,Kim2015,KimJerjen2015,Laevens2015a,Laevens2015b,
Bechtol2015,Koposov2015,Drlica-Wagner2015}).
Due to the relatively shallow depths of these surveys, however, UFDs at large distances
are undetectable and the total count of dwarf galaxies around the Milky Way
remains uncertain.

We are conducting an extensive search for UFDs using data from the Hyper Suprime-Cam
Subaru Strategic Program (HSC-SSP) \citep{Aihara2018a,Aihara2018b}.  HSC is a wide-field imager
installed at the prime focus of the Subaru Telescope \citep{Miyazaki2018}, and a 300-night
survey with this instrument is being carried out \citep{Aihara2018a}.
The combination of the superb image quality ($\sim 0''.6$) and depth ($i \sim 26$) and
the fact that it is a multi-band survey allows us to identify UFDs at large distances,
filling in the parameter space unexplored by previous surveys.
We have discovered two UFDs thus far, Virgo~I and
Cetus~III, as reported in \citet{Homma2016} and \citet{Homma2018}.  The current paper presents
the discovery of another UFD and a likely globular cluster from our continuing search.

The paper is structured as follows.  Section 2 describes our search method, followed
by detailed analyses of the identified stellar over-densities in Section 3.
Section 4 discusses implications of our results for dark matter models, and Section 5
concludes the paper.  Unless otherwise stated, magnitudes are given in the AB system.

\begin{figure*}[t!]
\begin{center}
\includegraphics[width=130mm]{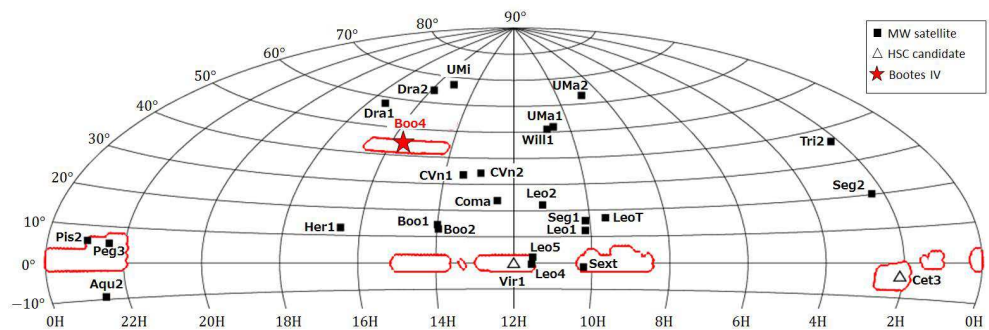}
\end{center}
\caption{
Survey areas of the HSC-SSP for S18A (bounded by red curves). A red star denotes the location of
the newly found satellite in this work, Bo\"otes IV. Filled squares and empty triangles denote,
respectively, the known satellites and those found in the HSC-SSP reported earlier.
We note that Pis~II and Leo~V are just outside the HSC-SSP footprint.
}
\label{fig: survey_area}
\end{figure*}

\section{Data and Method}

We make use of the imaging survey data from the HSC-SSP Wide layer, which plans to cover
$\sim 1,400$ deg$^2$ over the northern sky in five, $grizy$ photometric bands
(\cite{Aihara2018a,Aihara2018b,Furusawa2018,Kawanomoto2018,Komiyama2018,Miyazaki2018}).
The target 5$\sigma$ point-source limiting magnitudes in this Wide layer are given as
$g=26.5$, $r=26.1$, $i=25.9$, $z=25.1$ and $y=24.4$~mag.
In this paper, we adopt the $g$, $r$, and $i$-band data obtained
before 2018 April (internal data release S18A), which cover $\sim 676$ deg$^2$ in
6 separate fields along the celestial equator and one field around
$(\alpha_{\rm 2000},\delta_{\rm 2000})=(242^{\circ},43^{\circ})$
(Figure \ref{fig: survey_area}).
This is to be compared with $\sim 300$ deg$^2$ covered in our previous searches of
satellite galaxies from the data obtained before 2016 April \citep{Homma2018}.
The HSC data have been processed using
hscPipe v6.5 \citep{Bosch2018}, which is a branch of the Large Synoptic Survey Telescope
(LSST) pipeline
(see \cite{Ivezic2008,Axelrod2010,Juric2017}) calibrated against PS1 data of
photometry and astrometry (e.g., \cite{Tonry2012,Schlafly2012,Magnier2013}).
These photometry data in HSC-SSP are corrected for the mean foreground
extinction in the MW \citep{Schlafly2011}.

\subsection{Selection of stars}

We use the same method to select target stars as in our previous work.
In brief, we first select point-like images to avoid galaxies, set the criterion $g-r<1$ to remove
M-type main-sequence stars in the foreground, and then use the fiducial relation in the
$g-r$ vs. $r-i$ diagram anticipated for stars to eliminate other contaminants.

In order to extract point sources from the data, we use the {\it extendedness} parameter
calculated by the pipeline \citep{Homma2018}, which is defined on the basis of
the ratio between PSF and cmodel fluxes \citep{Abazajian2004}, $f_{\rm PSF}/f_{\rm cmodel}$.
A point source is defined to be an object with $f_{\rm PSF}/f_{\rm cmodel} > 0.985$.
In what follows, we make use of this parameter measured in the $i$-band, where the typical seeing
is the best of the five filters giving a median of $\sim 0''.6$.
\citet{Aihara2018b} presented,
by combining the data in HSC COSMOS with those in HST/ACS \citep{Leauthaud2007}, the definition
for the completeness and contamination in the star/galaxy classification.
The completeness is defined as the fraction of stars obtained by ACS which
are classified as stars correctly by HSC. This value is larger than 90\% at $i<22.5$, and
decreases to $\sim50\%$ at $i=24.5$.
The contamination is defined as the fraction of stars in the HSC classification
which are actually galaxies in ACS. This value is nearly zero at $i<23$ and increases to
$\sim 50\%$ at $i=24.5$. Here, we select stars as point sources down to $i=24.5$.

Next, we adopt the criteria for colors to remove the other objects such as disk stars,
background quasars and distant galaxies. These contaminants remained
even after the {\it extendedness} cut. We use the $g-r$ vs. $r-i$ color-color diagram
for stars with {\it extendedness}$=0$ and galaxies with {\it extendedness}$=1$
in the field called WIDE12H, where the bright limit of $i < 21$~mag is satisfied
so that the star/galaxy classification is robust.
Unlike galaxies, the distribution of stars in this color-color diagram follows a narrow
sequence (see Figure 1 of \citet{Homma2018}). Thus, following previous works (e.g., \cite{Willman2002}),
we select stars located inside this sequence, after removing red stars (dominated by M-type dwarfs
in the disk component) having a $g-r$ color of $g-r \ge 1$.
This star sequence is characterized by a polygon bounded by
($g-r$, $r-i$) of ($1.00$, $0.27$), ($1.00$, $0.57$), ($-0.4$, $-0.55$), and ($-0.4$, $-0.25$).
This color cut has a width of $\Delta (r-i) = 0.3$ mag, which is wider than twice the typical
1$\sigma$ photometric error of $r-i$ at $i = 24.5$. Thus, this method is optimal to select candidate
halo stars in the MW.

\subsection{Algorithm for the detection of the stellar overdensities}

Dwarf satellites in the MW are characterized as metal-poor, old stellar systems,
and they will present as statistically significant overdensities of stars
compared to the random distribution of MW field stars and distant galaxies/quasars.
We look for overdensities using the color-magnitude (CMD)
isochrone-based matched filter summarized below \citep{Homma2016,Homma2018}
following the approach of \citet{Koposov2008} and \citet{Walsh2009}.

\begin{figure}[t!]
\begin{center}
\includegraphics[width=85mm]{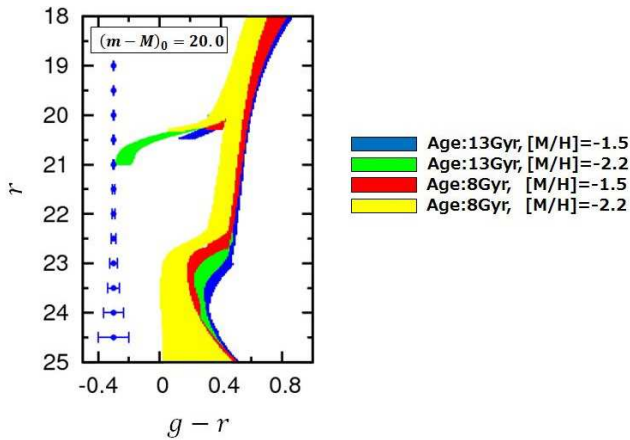}
\end{center}
\caption{
CMD filters at a distance modulus of $(m-M)_0 =20$~mag defined with an age of $t= 13$~Gyr and metallicity of
[M$/$H]$=-1.5$ (blue area), $t = 13$~Gyr and [M$/$H]$=-2.2$ (green area),
$t = 8$~Gyr and [M$/$H]$=-1.5$ (red area), and $t = 8$~Gyr and [M$/$H]$=-2.2$ (yellow area).
The error bars correspond to the typical error in the measurement of color at each $r$-band magnitude.
}
\label{fig: CMDfilter}
\end{figure}

\subsubsection{Isochrone filters}

The UFDs in the MW, that have been identified to date, show a similar
distribution in the CMD diagram to MW globular clusters, i.e., a characteristic
locus expected for metal-poor, old stellar systems. Following our previous work,
we thus select stars located inside a CMD locus for a metal-poor, old stellar system
within a specified distance interval from the Sun. This is a so-called isochrone filter,
for which we use a PARSEC isochrone \citep{Bressan2012} to define four different models:
(a) an age of $t =13$~Gyr and metallicity of [M$/$H]$=-1.5$,
(b) $t =13$~Gyr and [M$/$H]$=-2.2$,
(c) $t = 8$~Gyr and [M$/$H]$=-1.5$, and
(d) $t = 8$~Gyr and [M$/$H]$=-2.2$.

As a CMD, we use the $g-r$ color and $r$-band absolute magnitude, $M_r$. In order to
work with the current HSC filter system \citep{Kawanomoto2018}, the SDSS filter
system, which is available for the PARSEC isochrone, has been adopted and converted,
as given in the \citet{Homma2016} paper.
The finite selection width for the use of this isochrone filter is given as a function
of $r$-band magnitude. We adopt the quadrature sum of a 1$\sigma$ error in the measurement
of the $(g-r)$ color in HSC imaging and a color dispersion being typically $\sim \pm 0.05$ mag
expected for red giant-branch stars (RGBs) associated with a metallicity dispersion of
typically $\sim \pm 0.5$ dex as seen in MW dSphs.
This isochrone filter is shifted over distance moduli of $(m-M)_0 =16.5 - 24.0$
in steps of 0.5~mag, to search for metal-poor, old stellar systems at heliocentric distances of
$D_{\odot}=20 - 631$~kpc. In Figure \ref{fig: CMDfilter}, we show our four isochrone filters
placed at $(m-M)_0 = 20.0$.

\subsubsection{Method to identify overdensities and estimate their statistical significance}

Based on the isochrone filters at each distance from the Sun,
we search for spatial overdensities and estimate their statistical significance.
The stars selected in $0^{\circ}.05 \times 0^{\circ}.05$ boxes in right ascension
and declination are counted, where an overlap of $0^{\circ}.025$ in each direction is adopted.
We note that the grid interval of $0^{\circ}.05$ in this search is comparable to
a half-light diameter of $\sim 80$~pc for a typical UFD at $D_{\odot}=90$~kpc and that
dwarf satellites located at or beyond this characteristic distance, which are our targets
in HSC-SSP, can be identified within this grid interval.

In this procedure, we estimate the number of stars in each cell, $n_{i,j}$.
We ignore cells without any stars, $n_{i,j}=0$, which are for instance caused by
masking near a bright-star image. Then, for each of the separate Wide-layer fields,
we obtain the mean density ($\bar{n}$) and its dispersion ($\sigma$) over all cells 
and examine the normalized signal in each cell, $S_{i,j}$, which is the number of standard
deviations relative to the local mean \citep{Koposov2008,Walsh2009},
\begin{equation}
S_{i,j} = \frac{n_{i,j}-\bar{n}}{\sigma} \ .
\end{equation}
The distribution of $S$ is found to be almost Gaussian as obtained in our previous papers.

For the selection of candidate overdensities that have statistically high significance so that
false detections are removed, we set the detection threshold, $S_{\rm th}$,
for the value of $S$ \citep{Walsh2009}, following the result of a Monte Carlo analysis in
our previous work \citep{Homma2018}. The characteristic distribution of $S$ expected for
purely random fluctuations in stellar densities is obtained and we estimate $S_{\rm th}$
from a maximum density contrast found as a function
of $\bar{n}$. We adopt a threshold in the density contrast of $1.5 \times S_{\rm th}$ to
identify statistically significant overdensities, while avoiding contamination from random fluctuations.

\section{Results}

Using the algorithm above, we have detected two new overdensities beyond the above detection threshold,
in addition to the previously discovered UFDs in the survey footprint.
The first high signal is found at
$(\alpha_{\rm 2000},\delta_{\rm 2000}) = (15^{\rm h}34^{\rm m}45^{\rm s}.36, +43^{\circ}43'33''.6)$
in the direction of Bo\"otes
($S=32.3$ with $\bar{n}=1.48$, $S/S_{\rm th}=4.09$) and
the second one is found at
$(\alpha_{\rm 2000},\delta_{\rm 2000}) = (22^{\rm h}17^{\rm m}14^{\rm s}.16, +03^{\circ}28'48''.0)$
($S=31.8$ with $\bar{n}=1.60$, $S/S_{\rm th}=4.20$).
As explained below, the former is a dwarf galaxy candidate, hereafter called Bo\"otes~IV,
and the latter appears to be a compact globular cluster, hereafter called HSC~1.

\subsection{HSC~$J1534+4343$ - a new satellite, Bo\"otes~IV}

\begin{figure*}[t!]
\begin{center}
\includegraphics[width=120mm]{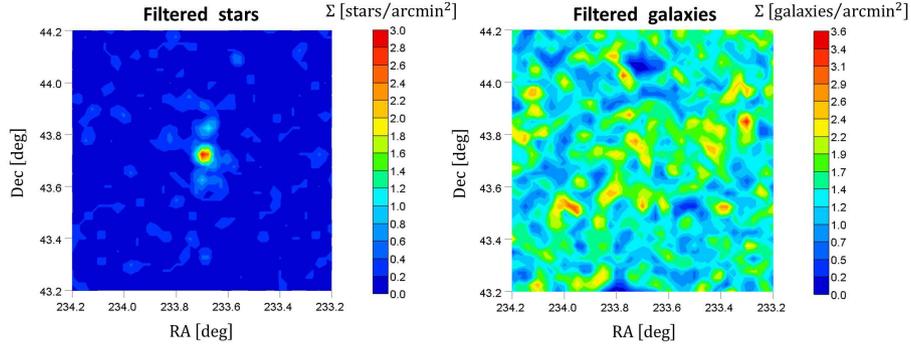}
\end{center}
\caption{
Left panel: the stellar overdensity found in Bo\"otes
passing the isochrone filter of $t=13$~Gyr and [M$/$H]$=-2.2$ at $(m-M)_0 = 21.6$
for the point sources satisfying $i < 24.5$, $g-r<1.0$, and the color-color cuts
expected for stars in the $g-r$ vs. $r-i$ diagram (see subsection 2.1)
The plot is shown over $1$~deg$^2$ centered on this UFD candidate.
Right panel: the spatial distribution of the galaxies which
pass the same isochrone filter and constraints as for the stars.
No overdensity is found in this plot.
}
\label{fig: Bootes_space}
\end{figure*}

This high-density stellar system is found through the isochrone filter of
$t=13$~Gyr and [M$/$H]$=-2.2$ placed at $(m-M)_0 = 21.6$~mag.
The left panel of Figure \ref{fig: Bootes_space} depicts this system, in which stars are inside
the isochrone filter at this distance modulus. The right panel corresponds to galaxies, which
do not show any overdensity.

\begin{figure*}[t!]
\begin{center}
\includegraphics[width=150mm]{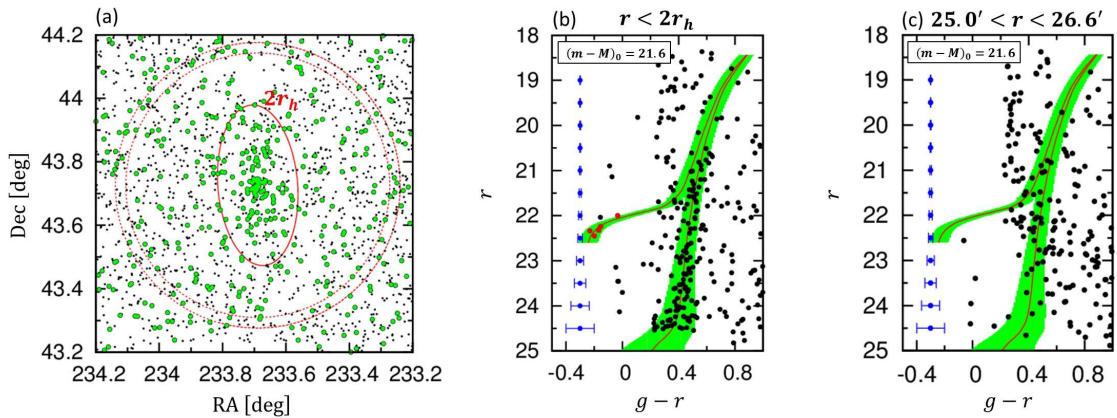}
\end{center}
\caption{
(a) The spatial distribution of the objects classified as stars around the overdensity in Bo\"otes.
The green circles and black dots denote, respectively. the stars inside and outside the isochrone filter
at the distance modulus of $(m-M)_0 = 21.6$.
A solid red curve shows an ellipse with a major axis of $r=2r_h$ ($r_h=7'.6$)
and an ellipticity of 0.64, whereas dashed red circles show annuli with radii $25'.0$ and $26'.6$
from the center of the overdensity.
(b) The CMD of the stars defined as the $g-r$ vs. $r$ relation, which are located within
the solid red ellipse in panel (a). 
Red circles denote the five BHB stars for the use of the additional distance estimate
as decribed in the text.
(c) The same as (b) but for field stars at $25'.0 < r < 26'.6$
having the same solid angle. A red giant-branch feature is absent in this plot.
}
\label{fig: Bootes_cmd}
\end{figure*}

In Figure \ref{fig: Bootes_cmd}(a), we plot the spatial distribution of the objects classified as stars
around the overdensity. There is a high-density region of stars within an ellipse with
a major axis of $r=2r_h$ ($r_h=7'.6$) and an ellipticity of 0.64, where these values are
obtained in Section 3.1.1.
Figure \ref{fig: Bootes_cmd}(b) shows the CMD defined with $(g-r, r)$ for the stars within
the ellipse given in panel (a). We identify a clear RGB locus, whereas
this feature is absent for the stars in the annulus with $25'.0 < r < 26'.6$ which has
the same solid angle, as shown in Figure \ref{fig: Bootes_cmd}(c).
These are probably field stars outside the overdensity.

\subsubsection{Structural parameters}

Using a likelihood analysis, we calculate the heliocentric distance to this stellar system,
for which we use RGB and blue horizontal-branch (BHB) stars 
inside the isochrone filter shown in Figure \ref{fig: Bootes_cmd}(b) obtained at
at the best-fit distance modulus of $(m-M)_0 = 21.6$ when we search for the overdensity.
If we assume a Gaussian distribution for the probability of the photometry of
these member stars relative to the best-fit CMD isochrone,
we estimate a best distance modulus of $(m-M)_0 = 21.6^{+0.2}_{-0.2}$, corresponding to
a heliocentric distance of $D_{\odot} = 209^{+20}_{-18}$~kpc.

As an alternative distance estimate, we identify the five blue horizontal-branch
(BHB) stars in the range of $22.0<r<22.5$ and $-0.3<g-r<0$ (red circles
in Figure \ref{fig: Bootes_cmd}(b)) and use these BHBs to determine
the distance following the formula for their absolute magnitude
calibrated by \citet{Deason2011}. The detailed description for this method
is given in \citet{Fukushima2019}. We find that
this procedure provides $(m-M)_0 = 21.59^{+0.14}_{-0.14}$, taking into account
the typical photometric error in $g-r$ (0.02 mag) for the $r$ magnitude of these BHBs.
Since this estimate is consistent with the above distance estimate using the isochrone
filter, we adopt $(m-M)_0 = 21.6^{+0.2}_{-0.2}$ in what follows.

Following \citet{Martin2008} and \citet{Martin2016}, we obtain the structural properties of
this overdensity. We define six parameters $(\alpha_0, \delta_0, \theta, \epsilon,
r_h, N_{\ast})$, where each of the parameters denotes the following meaning.
$(\alpha_0, \delta_0)$ denote the celestial coordinates
for the center of the overdensity, $\theta$ is its position angle from North to
East, $\epsilon$ is the ellipticity, $r_h$ is the half-light radius measured
on the major axis, and $N_{\ast}$ is the number of stars within the isochrone
and brighter than our magnitude limit, which belong to the overdensity. 
We then perform the maximum likelihood analysis given in \citet{Martin2008} for the stars
within a circle of radius $30'$, corresponding to $\sim 3.9$ times the derived $r_h$,
which pass the above isochrone filter. We summarize the results in Table~\ref{tab: 1}.

\begin{table}
\tbl{Properties of Bo\"otes~IV}{
\begin{tabular}{lc}
\hline
Parameter$^{a}$ & Value \\
\hline
Coordinates (J2000)           & $233^{\circ}.689$, $43^{\circ}.726$      \\
Galactic Coordinates ($l,b$)  &  $70^{\circ}.682$, $53^{\circ}.305$      \\
Position angle	              & $+3^{+4}_{-4}$ deg        \\
Ellipticity                   & $0.64^{+0.05}_{-0.05}$    \\
Number of stars, $N_{\ast}$   & $124^{+10}_{-10}$         \\
Extinction, $A_V$             & 0.067 mag                 \\
$(m-M)_0$                     & $21.6^{+0.2}_{-0.2}$ mag  \\
Heliocentric distance         & $209^{+20}_{-18}$ kpc     \\
Half light radius, $r_h$      & $7'.6^{+0'.8}_{-0'.8}$ or $462^{+98}_{-84}$ pc \\
$M_{{\rm tot},V}$             & $-4.53^{+0.23}_{-0.21}$ mag \\
\hline
\end{tabular}}\label{tab: 1}
\begin{tabnote}
$^{a}$Integrated magnitudes are corrected for the mean Galactic foreground extinction,
$A_V$ \citep{Schlafly2011}.
\end{tabnote}
\end{table}

\begin{figure}[t!]
\begin{center}
\includegraphics[width=80mm]{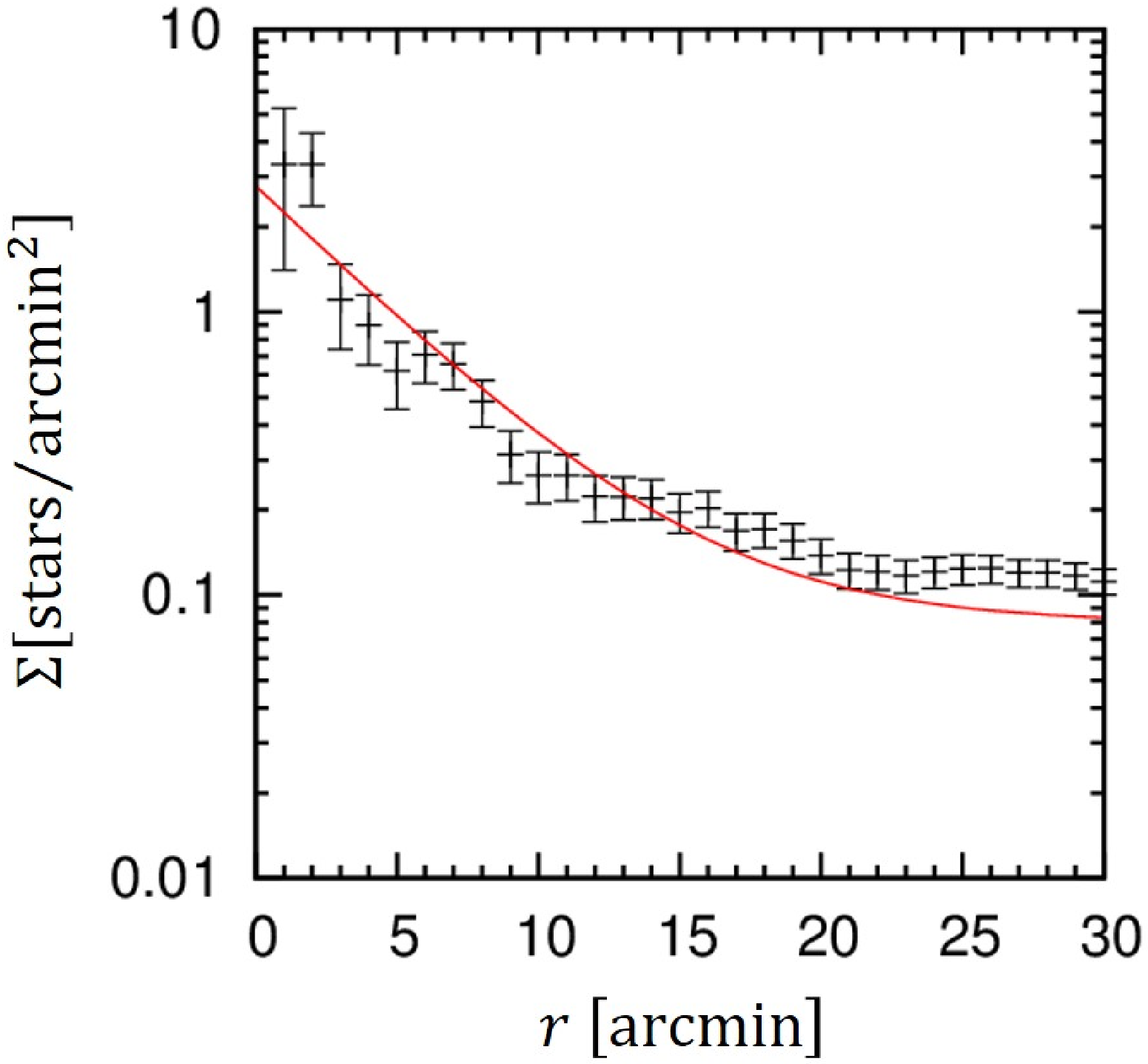}
\end{center}
\caption{The density distribution of the stars in Bo\"otes~IV passing the isochrone filter
given in Figure \ref{fig: Bootes_cmd}(b), in elliptical annuli as a function of mean radius.
The uncertainties are based on Poisson statistics.
The red line denotes the best-fit exponential profile with $r_h = 7'.6$, plus a constant
representing the background.}
\label{fig: Bootes_profile}
\end{figure}

We show, in Figure \ref{fig: Bootes_profile}, the radial distribution of the stars
passing the isochrone filter depicted in Figure \ref{fig: Bootes_cmd}(b), which
is obtained from the count of the average density within
elliptical annuli. The red line in the figure denotes the best-fit
exponential profile having $r_h = 7'.6$ or 462~pc.
We note that this spatial size is quite large compared with the typical scale of
MW globular clusters, but is in agreement with that of MW dwarf galaxies,
as we describe in Section 4.3.

\subsubsection{$V$-band absolute magnitude}

Here we estimate $M_V$ of Bo\"otes~IV in the following manner. We first
transform $(g, r)$ to $V$ using the formula given in \citet{Jordi2006}. which are
calibrated for metal-poor, Population II stars.
We then perform a Monte Carlo procedure similar to \citet{Martin2008} for the
estimation of the most likely value of $M_V$ and its uncertainty. 
Using the values of $N_{\ast} = 124^{+10}_{-10}$ at $i < 24.5$ mag and
$(m-M)_0 = 21.6^{+0.2}_{-0.2}$ mag as obtained in the procedure of Section 3.1.1
as well as a model for the stellar population having an age of 13~Gyr and
metallicity of [M$/$H]$=-2.2$, the $10^4$ realizations of CMDs are generated
for the initial mass function (IMF) given by \citet{Kroupa2002}.
The luminosity of the stars at $i < 24.5$ mag is then derived, also considering
the completeness of the stars observed with HSC. We obtain
the median value and the 68\% confidence interval of $M_V$
as $M_V = -4.53^{+0.23}_{-0.21}$~mag. These values are
very insensitive to whether we choose the Salpeter \citep{Salpeter1955} or
Chabrier IMF \citep{Chabrier2001}.
The total mass of stars in Bo\"otes~IV is also estimated for different IMFs, as
$M_{\ast}=14430^{+3350}_{-3939} M_{\odot}$ (Kroupa),
$13779^{+3161}_{-3693} M_{\odot}$ (Chabrier), and
$27730^{+6477}_{-7732} M_{\odot}$ (Salpeter).


\subsection{HSC~$J2217+0328$ - a compact star cluster candidate, HSC~1}

\begin{figure*}[t!]
\begin{center}
\includegraphics[width=120mm]{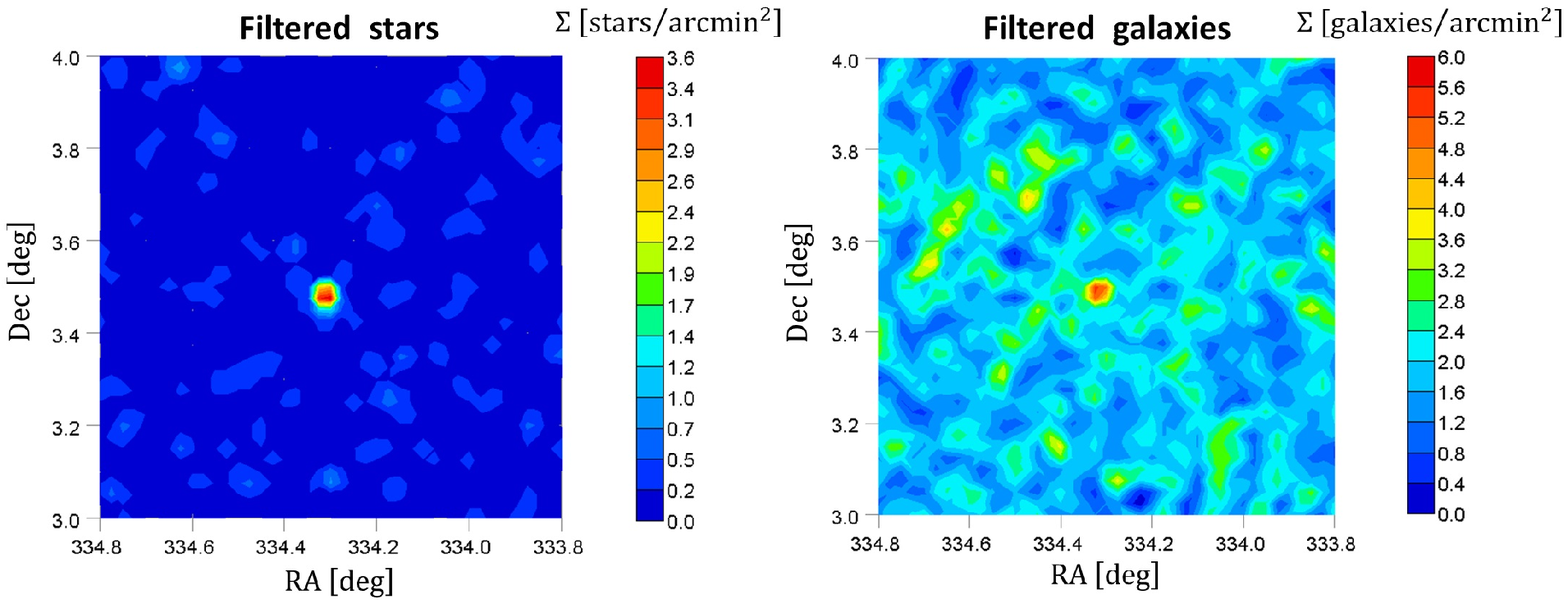}
\end{center}
\caption{
Left panel: the stellar overdensity found in the vicinity of a star cluster
candidate, HSC~1, passing the isochrone filter of $t=13$~Gyr and [M$/$H]$=-2.2$ at $(m-M)_0 = 18.3$
for the point sources satisfying $i < 24.5$, $g-r<1.0$, and the color-color cuts
expected for stars in the $g-r$ vs. $r-i$ diagram (see subsection 2.1).
The plot is shown over $1$~deg$^2$ centered on this star cluster candidate.
Right panel: the spatial distribution of the galaxies
which pass the same isochrone filter and constraints as for the stars.
An overdensity at the center of this plot is seen, possibly because of
misclassification of faint stars within a cluster as galaxies in hscPipe.
}
\label{fig: Cluster_space}
\end{figure*}

\begin{figure}[t!]
\begin{center}
\includegraphics[width=60mm]{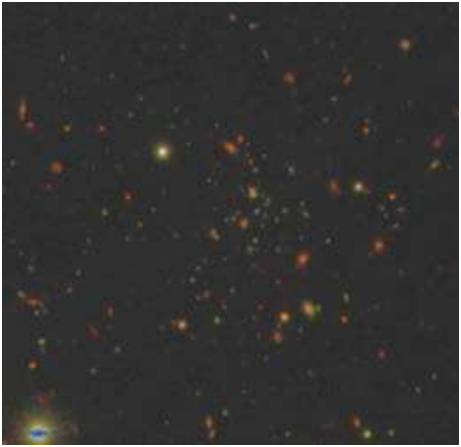}
\end{center}
\caption{
The $gri$ color map of the HSC image within a 2' x 2' box around the overdensity, HSC~1.
}
\label{fig: Cluster_image}
\end{figure}

We have detected this overdensity having $S/S_{\rm th}=4.20$ and $\bar{n}=1.60$ (31.8$\sigma$)
is found with the isochrone filter of
$t=13$~Gyr and [M$/$H]$=-2.2$ placed at
$(\alpha_{\rm 2000}, \delta_{\rm 2000})=(334^{\circ}.309, 3^{\circ}.480)$ and $(m-M)_0 = 18.3$~mag.
We show, in Figure \ref{fig: Cluster_space}, this overdensity passing the isochrone filter
at this distance for the objects classified as stars (left) and galaxies (right).
While there is a clear overdensity signal in stars, we also identify an overdensity in galaxies
at the same position. We suggest that this is caused by the misclassification of faint stars
within a cluster as galaxies in hscPipe.
To get insights into this overdensity feature, we plot the $gri$ color map of the HSC image
around its location in Figure \ref{fig: Cluster_image}. One can see the concentration
of point sources directly by eye.
This overdensity is a candidate star cluster, which is hereafter called HSC~1
because it is the first new cluster found with HSC.

\begin{figure*}[t!]
\begin{center}
\includegraphics[width=150mm]{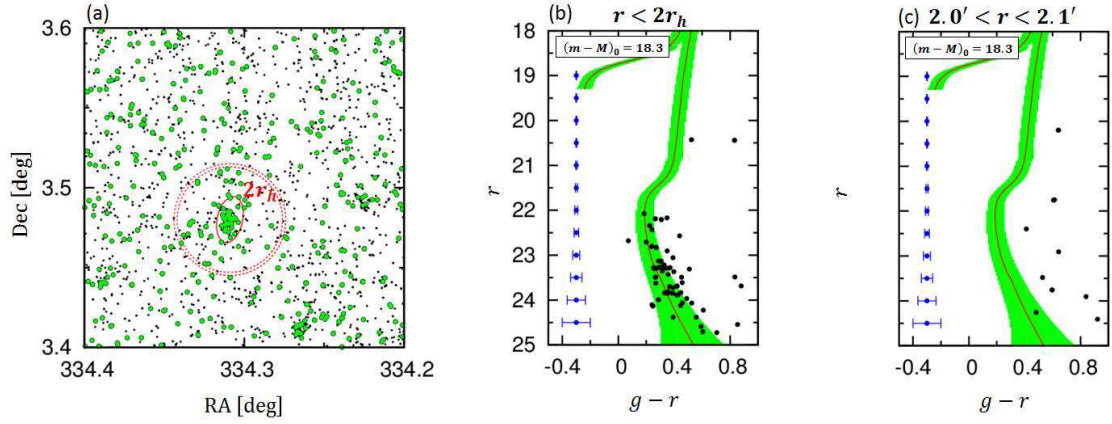}
\end{center}
\caption{
(a) The spatial distribution of the objects around the overdensity, HSC~1.
The green circles and black dots denote, respectively, the objects inside and outside the isochrone filter
at $(m-M)_0 = 18.3$. In this plot, all of the objects without taking into account
the {\it extendedness} parameter are shown to avoid the effect of the deblender failure
in the star/galaxy separation.
A solid red curve shows an ellipse with a major axis of $r=2r_h$ ($r_h=0'.44$)
and an ellipticity of 0.46, whereas dashed red circles mark an annulus with radii $2'.0$ and $2'.1$
from the center.
(b) The CMD of the objects defined in the $g-r$ vs. $r$ CMD relation, which are located
within the solid red ellipse in panel (a), where we use the undeblended flux of the HSC image
with a seeing-matched aperture magnitude to avoid the effect of the deblender failure
in this crowded region.
(c) The same as (b) but for the objects at $2'.0 < r < 2'.1$ having
the same solid angle, i.e., the field objects outside the overdensity.
Note the absence of a main sequence feature seen in (b).
}
\label{fig: Cluster_cmd}
\end{figure*}

As mentioned above, hscPipe has misclassified some stars as galaxies within the cluster,
which may be due to the deblender failure of this crowded region and the resultant failure of
the star/galaxy separation. To mitigate this effect, we use the undeblended flux of the HSC image
with a seeing-matched aperture magnitude to select likely member objects belonging to this overdensity.
In Figure \ref{fig: Cluster_cmd}(a), we plot the spatial distribution of all objects after
color cuts around this overdensity, both those classified as stars and galaxies.
There is a localized concentration of objects within an ellipse
with a semi-major axis of $r_h=0'.44$ and an ellipticity of 0.46.
Figure \ref{fig: Cluster_cmd}(b) shows the CMD defined with $(g-r, r)$ for objects
within the solid red ellipse shown in panel (a).
We have found a clear signature of a main sequence (MS), whereas
there is no such feature for the objects at $2'.0 < r < 2'.1$ having the same solid angle,
as shown in Figure \ref{fig: Cluster_cmd}(c).
These are probably field stars outside the overdensity.

In the same way as for Bo\"otes~IV, we estimate heliocentric distance, structural parameters
and V-band absolute magnitude for this stellar system as follows. We adopt PSF magnitudes
for these estimates.

Based on a likelihood analysis, we estimate the $D_{\odot}$ to this stellar system,
for which we use MS stars inside the isochrone filter shown in Figure \ref{fig: Cluster_cmd}(b)
obtained at the best-fit distance modulus of $(m-M)_0 = 18.3$ when we search for the overdensity.
We obtain a best distance of $(m-M)_0 = 18.3^{+0.2}_{-0.2}$, corresponding to
a heliocentric distance of $D_{\odot} = 46^{+4}_{-4}$~kpc.

The structural properties of this stellar system are derived by applying the maximum likelihood method
to the objects \citep{Martin2008} located within a circle of radius $3'.0$
passing the isochrone filter, which corresponds to $\sim 6.8$ times the anticipated $r_h$.
We note that this method models the contribution for the background uniformly distributed objects
like galaxies, and therefore is applicable to the structural analysis of the star cluster
even though we have not filtered out the galaxies.
The results are summarized in Table~\ref{tab: 2}.

\begin{table}
\tbl{Properties of a star cluster candidate, HSC~1}{
\begin{tabular}{lc}
\hline
Parameter$^{a}$ & Value \\
\hline
Coordinates (J2000)           & $334^{\circ}.309$, $3^{\circ}.480$      \\
Galactic Coordinates ($l,b$)  &  $66^{\circ}.319$, $-41^{\circ}.841$      \\
Position angle	              & $-12^{+11}_{-11}$ deg        \\
Ellipticity                   & $0.46^{+0.08}_{-0.10}$    \\
Number of stars, $N_{\ast}$   & $47^{+6}_{-6}$            \\
Extinction, $A_V$             & 0.222 mag                 \\
$(m-M)_0$                     & $18.3^{+0.2}_{-0.2}$ mag  \\
Heliocentric distance         & $46^{+4}_{-4}$ kpc     \\
Half light radius, $r_h$      & $0'.44^{+0'.07}_{-0'.06}$ or $5.9^{+1.5}_{-1.3}$ pc \\
$M_{{\rm tot},V}$             & $-0.20^{+0.59}_{-0.83}$ mag \\
\hline
\end{tabular}}\label{tab: 2}
\begin{tabnote}
$^{a}$Integrated magnitudes are corrected for the mean Galactic foreground extinction,
$A_V$ \citep{Schlafly2011}.
\end{tabnote}
\end{table}

\begin{figure}[t!]
\begin{center}
\includegraphics[width=80mm]{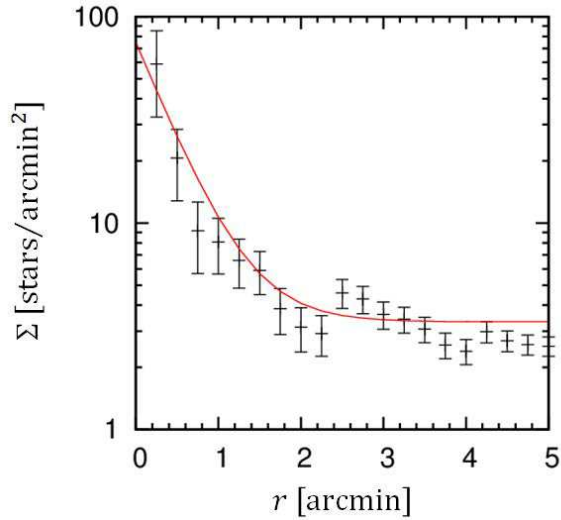}
\end{center}
\caption{The density distribution of the stars in HSC~1 passing the isochrone filter
given in Figure \ref{fig: Cluster_cmd}(b), in elliptical annuli as a function of mean radius.
The uncertainties are based on Poisson statistics.
The red line denotes the best-fit exponential profile with $r_h = 0'.44$.}
\label{fig: Cluster_profile}
\end{figure}

Figure \ref{fig: Cluster_profile} shows the radial density distribution of the stars
passing the isochrone filter, which is obtained from the count of the average density within
elliptical annuli. The red line denotes the best-fit
exponential profile having $r_h = 0'.44$ or 5.9~pc.

The $M_V$ of this stellar system is estimated in the same way
as for Bo\"otes~IV and we obtain $M_V = -0.20^{+0.59}_{-0.83}$~mag.
We also calculate the total mass of stars in HSC~1 for different IMFs as
$M_{\ast}=254^{+58}_{-51} M_{\odot}$ (Kroupa), $255^{+55}_{-51} M_{\odot}$ (Chabrier), and
$487^{+111}_{-99} M_{\odot}$ (Salpeter).
The combination of the absolute magnitude and small half-light radius suggests that
HSC~1 is a compact star cluster as will be discussed in Section 4.1. That is, this is a
MW globular cluster as judged from its location far beyond the MW disk.
However, to really distinguish between star clusters and galaxies, we need to determine
if this object is embedded in a dark matter halo, which will require measuring
accurate stellar radial velocities.

\section{Discussion}

\subsection{Comparison with MW globular clusters and dwarf galaxies}

We first compare the structural properties of Bo\"otes~IV and HSC~1 with those of previously 
known MW globular clusters and dwarf satellites. Globular clusters are characterized by
their compact size and round shape without having a surrounding dark halo, suggesting
a self-gravitating stellar system. On the other hand, dwarf satellites have an extended
spatial distribution of stars, and their velocity distribution
strongly suggests the presence of a surrounding dark halo.

Figure \ref{fig: abs_size}(a) shows the relation between the half-light radii and
absolute magnitudes of globular clusters in the MW (dots) \citep{Harris1996} and those of
dwarf satellites (squares) \citep{McConnachie2012,Laevens2014,Bechtol2015,
Koposov2015,Drlica-Wagner2015,Kim2015,KimJerjen2015,Laevens2015a,Laevens2015b,
Torrealba2016,Torrealba2018}.
Red and blue stars with error bars, respectively, denote Bo\"otes~IV and HSC~1.
It is clear that Bo\"otes~IV is significantly large compared with MW globular clusters
having similar $M_V$ and that it follows the locus of MW dwarf satellites.
This suggests that Bo\"otes~IV is a typical UFD, whose shape is flattened with an ellipticity of
$\epsilon=0.64^{+0.05}_{-0.05}$. Globular clusters have much smaller ellipticities
at its brightness \citep{Harris1996,Marchi-Lasch2019}, while some dwarf galaxies have
even larger ellipticities, e.g., UMa~I having $\epsilon=0.80$ \citep{Martin2008},
supporting this conclusion.
On the other hand, the size of HSC~1 is comparable to MW globular clusters.
This suggests that HSC~1 is a globular cluster, although it is still possible that it is
a compact dwarf galaxy because HSC~1 lies on the $M_V$ vs. $r_h$ relation for dwarf galaxies
and its ellipticity, $\epsilon=0.46$, at $M_V = -0.20$ mag is typical for both dwarf galaxies
and globular clusters at such faint magnitudes \citep{Marchi-Lasch2019}.

\subsection{Detectability of Bo\"otes~IV and HSC~1 in other surveys}

The footprint of HSC-SSP overlaps that of SDSS, so we investigate the detectability
of Bo\"otes~IV and HSC~1 in both surveys.
For this purpose, we evaluate the completeness distance, $R_{\rm comp}^{\rm SDSS}$,
derived for SDSS, above which a satellite with a particular luminosity is undetectable, and
the corresponding distance for HSC-SSP, $R_{\rm comp}^{\rm HSC}$.
Following \citet{Koposov2008}, the former is given as,
\begin{equation}
R_{\rm comp}^{\rm SDSS} = 10^{(-a^{\ast} M_V - b^{\ast})} \ {\rm Mpc} \ ,
\label{eq: Rcomp_SDSS}
\end{equation}
where $(a^{\ast}, b^{\ast}) = (0.205, 1.72)$.
$R_{\rm comp}^{\rm HSC}$ is then given as \citep{Tollerud2008},
\begin{equation}
R_{\rm comp}^{\rm HSC}/R_{\rm comp}^{\rm SDSS} =
                      10^{(M_{r,{\rm HSC}} - M_{r,{\rm SDSS}})/5}  \ ,
\label{eq: Rcomp_HSC}
\end{equation}
where $M_{r,{\rm HSC}}$ and $M_{r,{\rm SDSS}}$ are, respectively, the limiting point-source
magnitudes in $r$-band for HSC and SDSS. In this paper,
we adopt $M_{r,{\rm SDSS}}=22.2$~mag at $\sim 100$\%
completeness and $M_{r,{\rm HSC}}=24.7$~mag at $\sim 50$\% completeness,
because the cut-off $i$-band magnitude adopted in this work is set as 
$i = 24.5$~mag and the stars in MW satellites typically have $r-i=0.2$.

\begin{figure}[t!]
\begin{center}
\includegraphics[width=85mm]{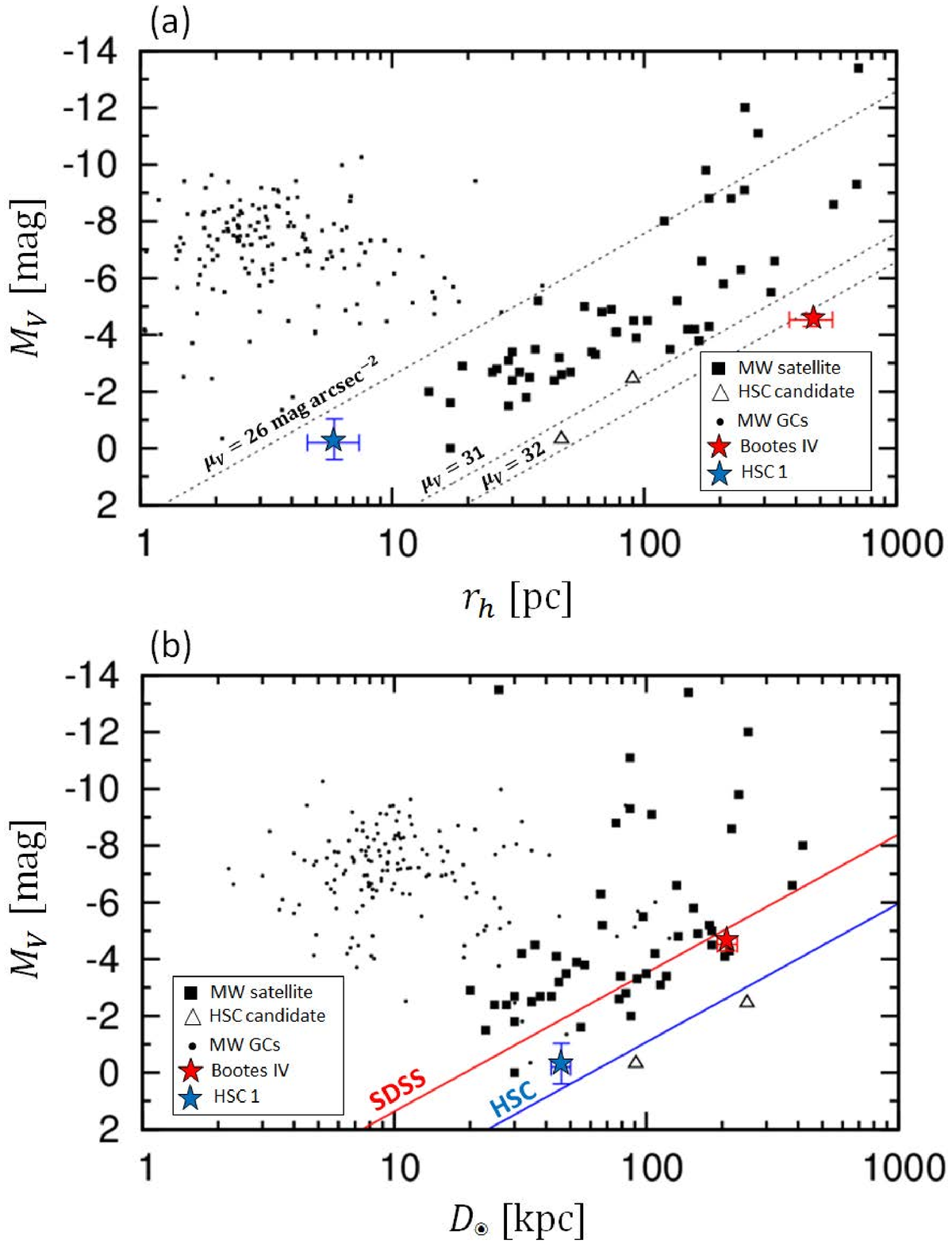}
\end{center}
\caption{(a) The relation between $M_V$ and $r_h$ for MW globular clusters (dots)
taken from \citet{Harris1996} and MW dSphs (filled squares) from \citet{McConnachie2012,
Laevens2014,Bechtol2015,Koposov2015,Drlica-Wagner2015,Kim2015,KimJerjen2015,
Laevens2015a,Laevens2015b,Torrealba2016,Torrealba2018}.
The empty triangles denote those found from the
previous data release of the HSC-SSP \citep{Homma2016,Homma2018}.
The red and blue stars with error bars, respectively, denote Bo\"otes~IV and HSC~1.
The dotted lines denote the loci of constant surface brightness, $\mu_V = 26$, 31,
and 32 mag~arcsec$^{-2}$.
(b) The relation between $M_V$ and heliocentric distance, $D_{\odot}$.
The red and blue lines denote, respectively, the detection limits of SDSS and HSC.
}
\label{fig: abs_size}
\end{figure}

Figure \ref{fig: abs_size}(b) presents the $M_V$ vs. $D_{\odot}$ relation
for MW globular clusters and dwarf galaxies including Bo\"otes~IV and HSC~1.
Red and blue lines show, respectively, the detection limits of SDSS and HSC evaluated
from equations (\ref{eq: Rcomp_SDSS}) and (\ref{eq: Rcomp_HSC}).
As is clear, both Bo\"otes~IV and HSC~1 are beyond the reach of SDSS but
(not surprisingly) within the limit of HSC.

\subsection{Implication for dark matter models}

We have discovered one new dwarf satellite, Bo\"otes~IV, from the HSC-SSP data released internally
in the outer halo of the MW, in addition to two satellites, Virgo~I and Cetus~III,
which we discovered in the previous data release \citep{Homma2016,Homma2018}.
To compare this discovery rate with the prediction of $\Lambda$CDM models,
we adopt the recent study of \citet{Newton2018} for the number of visible satellites
expected in $\Lambda$CDM models.

\citet{Newton2018} combined the data from SDSS and DES to infer the full complement of MW satellite
galaxies. For this purpose, they adopted the subhalo populations for MW-sized halos performed by
the simulation project, AQUARIUS \citep{Springel2008},
to obtain a prior for the radial distribution of satellites.
They made use of dark-matter-only simulations for six MW-sized halos and considered two effects
that affect the radial distribution of subhalos at redshift $z=0$. The first is due to
the finite resolution of the simulation, which leads to a number of spatially unresolved subhalos. 
They made use of a semi-analytical galaxy formation model to account for these otherwise unresolved
subhalos (or 'orphan galaxies'). Second, they considered the effect of subhalo depletion due to
tidal interaction with a central baryonic disk (an effect which is not included in this pure DM
simulation). These two effects work in opposite senses, and they found that the normalized radial
distribution of these fiducial subhalos is well fit to a so-called Einasto profile
\citep{Einasto1965, Navarro2004} and is similar to that of luminous satellites
obtained from high-resolution hydrodynamic simulations from the project called APOSTLE
\citep{Fattahi2016,Sawala2016b}.
For an assumed MW halo mass of $1.0 \times 10^{12}$ $M_{\odot}$, they find that
the total number of satellites with $M_V \le 0$ within 300~kpc of
the Sun is $124^{+40}_{-27}$ (see their Table ~E1); this number is only weakly
dependent on halo mass. This cumulative luminosity function is denoted as $N (M_V)$
in what follows.

Given this predicted number of visible satellites,
we evaluate the HSC-SSP survey area and depth to obtain
the actual number of satellites it should be able to detect.
First, HSC-SSP in its Wide layer has surveyed over $\sim 676$~deg$^2$, which corresponds to
a fraction of the sky, $f_{\Omega,{\rm HSC}} = 0.016$.
The number of expected satellites over this solid angle is given by
$f_{\Omega,{\rm HSC}} N (M_V)$.

Second, we evaluate the completeness correction associated with the detection limit of HSC,
$f_{r,{\rm HSC}}$, which depends on $M_V$. Following \citet{Newton2018}, we assume that
the radial distribution of satellites in this case follows an Einasto profile and
is independent of their luminosities. With this, $f_{r,{\rm HSC}}$ is given by
\begin{equation}
f_{r,{\rm HSC}} = \frac{N(<R_{\rm comp}^{\rm HSC})}{N(<300~{\rm kpc})} =
\frac{\gamma \left( \frac{3}{\alpha},
        \frac{2}{\alpha}\left[ c_{200}\frac{R_{\rm comp}^{\rm HSC}}{R_{200}} \right]^{\alpha} \right)
    }{\gamma \left( \frac{3}{\alpha},
        \frac{2}{\alpha}\left[ c_{200}\frac{300~{\rm kpc}}{R_{200}} \right]^{\alpha} \right) } \ ,
\end{equation}
where $\gamma$ is the lower incomplete Gamma function, defined as
\begin{equation}
\gamma(s,x) =  \int_0^x t^{s-1} \exp(-t) dt \ .
\end{equation}
The Einasto profile,
$\ln N(r)/N_{-2} = -(2/\alpha) [(r/r_{-2})^{\alpha}-1]$,
is parameterized by a shape parameter $\alpha$ and the concentration,
$c_{200}=R_{200}/r_{-2}$, where $r_{-2}$ denotes the radius at which the logarithmic slope of
the profile is $-2$ and the density there is $N_{-2}$.
We adopt $\alpha = 0.24$, $c_{200}=4.9$ and $R_{200} = 200$~kpc for
a $1.0 \times 10^{12}$ $M_{\odot}$ MW-sized halo \citep{Newton2018}.
The total number of satellites predicted for the current HSC-SSP data, after the combined
correction for both of the above effects, is now given as
$f_{r,{\rm HSC}} f_{\Omega,{\rm HSC}} N(M_V)$.

\begin{figure}[t!]
\begin{center}
\includegraphics[width=85mm]{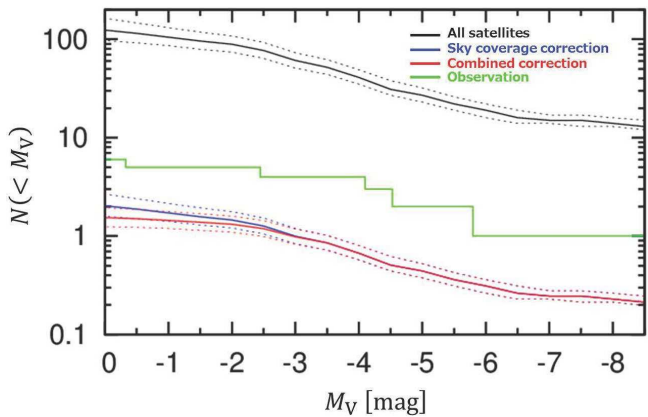}
\end{center}
\caption{
The black solid line shows the cumulative luminosity distribution of all visible satellites
obtained from the study of \citet{Newton2018}, namely
$124^{+40}_{-27}$ for $M_V \le 0$.
The blue solid line accounts for the correction related to the sky coverage only,
$f_{\Omega,{\rm HSC}}$, of the HSC-SSP survey and the red solid line accounts for
all the corrections associated with the HSC-SSP survey.
Dotted lines above and below each of the black,
blue and red solid lines correspond to the 68~\% confidence range
associated with these estimates.
The green solid line corresponds to the observed cumulative luminosity distribution of 
satellites identified within the HSC-SSP footprint of $\sim 676$~deg$^2$, which consists
of Sextans, Leo~IV, Pegasus~III, Cetus~III, Virgo~I and Bo\"otes~IV.
}
\label{fig: lumifun}
\end{figure}

In Figure \ref{fig: lumifun}, the black solid line shows the predicted
cumulative luminosity function of all the visible satellites
from the study of \citet{Newton2018}, namely $124^{+40}_{-27}$ for $M_V \le 0$.
The blue solid line denotes the luminosity function considering the correction for
the sky coverage only, $f_{\Omega,{\rm HSC}}$, whereas the
red solid line includes the full correction, $f_{r,{\rm HSC}} f_{\Omega,{\rm HSC}}$,
i.e., the expected number of visible satellites in the HSC-SSP survey of S18A,
given as $1.5^{+0.4}_{-0.3}$ for $M_V \le 0$.
We note that this number is very weakly dependent on the values for
$(a^{\ast},b^{\ast})$ in equation (\ref{eq: Rcomp_SDSS}); varying these over a reasonable
range changes this number by only $\sim 0.1$.
In addition, \citet{Newton2018} find that the total number of satellites is insensitive to
the assumed MW halo mass;
varying that mass from $0.5$ to $2.0 \times 10^{12}$ $M_{\odot}$ changes the cumulative number, $N$,
by no more than the 68\% uncertainty range indicated above.

The green solid line in Figure \ref{fig: lumifun} corresponds to the cumulative luminosity
distribution of the identified satellites within the HSC-SSP footprint of $\sim 676$~deg$^2$.
This sample includes six satellites of 
Sextans ($M_V=-9.3$: classical dwarf), Leo~IV ($M_V=-5.8$: SDSS DR9), Pegasus~III ($M_V=-3.4$: SDSS DR9), 
Cetus~III ($M_V=-2.4$: HSC), Virgo~I ($M_V=-0.3$: HSC) and Bo\"otes~IV.
That is, we find six dwarf galaxies, whereas the model predicted only $1.5^{+0.4}_{-0.3}$.
Thus, we apparently have a problem of too many satellites, instead of a missing satellites problem.
We note that the \citet{Newton2018} prediction is normalized with the sample of visible
satellites found in SDSS and DES, so it is not sensitive to the information of satellites
at much larger distances as those found by the current deeper survey. Also, 
the larger number of observed satellites may be due to the effect of a massive dwarf galaxy
with $10^9$ to $10^{10}$ $M_{\odot}$ as one of main progenitors for the formation of the stellar halo,
which merged with the MW around 6 to 10~Gyr ago,
as suggested from the kinematics analysis of nearby
stars from {\it Gaia} (e.g., \cite{Fattahi2018,Belokurov2018,Helmi2018}).
If that is the case, then this progenitor would bring additional subhalos and thus visible
satellites with it, as suggested from the discovery of UFDs associated with the Magellanic Clouds
in the DES survey \citep{Jethwa2016,Dooley2017}, as first suggested by \citet{Lynden-Bell1976}.
Also, there may exist some diversity in the luminosity function of satellites in MW-sized host
galaxies, as suggested from the recent study of nearby galaxies outside the Local Group \citep{Tanaka2018},
implying that the total population of subhalos is actually sensitive to the characteristics of
each MW-sized host halo, including their total mass and merging history.
Indeed, if we adopt the abundance matching model by \citet{Dooley2017} as adopted in
our previous papers, the predicted number of satellites in the HSC-SSP survey area is estimated as
$N=4^{+8}_{-2}$, i.e., in agreement with the current discovery rate of satellites
within a $1\sigma$ uncertainty.
This may suggest that the models need more refinements for the assignment of dark subhalos to
visible satellites.

The answer to this issue is beyond the scope of this work, but will be available as
further UFDs are found in the on-going HSC-SSP and
the LSST \citep{Abell2009}.

\section{Conclusions}

From the HSC-SSP data obtained through 2018 April,
we have found a highly compelling UFD candidate, Bo\"otes~IV,
and a globular cluster candidate, HSC~1. These objects were identified as
statistically high overdensities of 32.3$\sigma$ and 31.8$\sigma$, respectively,
in the relevant isochrone filter with age of 13~Gyr and metallicity of [M$/$H]$=-2.2$.
Based on a maximum likelihood analysis, the half-light radius of Bo\"otes~IV
is obtained as $r_h = 462^{+98}_{-84}$ pc and its V-band absolute magnitude is
$M_V = -4.53^{+0.23}_{-0.21}$~mag. This follows the size-brightness relation for other MW
dwarf satellites, suggesting that it is a dwarf galaxy.
For HSC~1, we have obtained $r_h = 5.9^{+1.5}_{-1.3}$~pc and $M_V = -0.20^{+0.59}_{-0.83}$~mag,
thus suggesting that it is a globular cluster because its size is comparable with
MW globular clusters having the same luminosity, although spectroscopic
studies are needed to determine if there is evidence for associated dark matter.

In the $\sim 676$ deg$^2$ covered to date in $gri$ in the HSC-SSP footprint, we have identified
three new UFDs from the HSC data (Virgo~I, Cetus~III and Bo\"otes~IV) and there exist
three previously known satellites.
To investigate what this number of satellites implies in view of $\Lambda$CDM models,
we adopt the recent theoretical estimate of satellite populations in a MW-sized halo
by \citet{Newton2018}, in which the sample of luminous satellites found in SDSS and DES are
considered and a prior for their radial distribution is inferred from
the AQUARIUS suite of high-resolution dark-matter only simulations, taking into account
both the effects of the unresolved population of subhalos due to finite numerical resolution and
their depletion due to tidal forces from the central disk galaxy. This model
predicts that HSC should have seen $1.5^{+0.4}_{-0.3}$ satellites with $M_V \le 0$ within
the HSC detection limit, to be compared with the observed number of six.
The \citet{Newton2018} prediction is insensitive to the mass of
the MW halo and the adopted values for the parameters relevant for identifying satellites.
However, the fact that we observe more satellites than predicted may be due to the presence of
diversity in the total satellite population depending on the characteristics of each
MW-sized host halo including their total mass and merging history.
For instance, analysis of the {\it Gaia} sample of nearby stars \citep{Belokurov2018,Helmi2018}
indicates that a $10^9-10^{10}$ $M_{\odot}$ galaxy fell into
the MW stellar halo as suggested from their possible debris present in the sample;
this Large Magellanic Cloud-class satellite may be associated with additional subhalos
as suggested by analyses of the DES survey \citep{Jethwa2016,Dooley2017}.
Another possibility is that the \citet{Newton2018} model is not
sensitive to the information of more distant satellites as found by the current deeper survey
than SDSS and DES. This suggests that the theoretical models need more refinements
for the assignment of dark subhalos to luminous satellites.

Before concluding this, we require spectroscopic follow-up studies of UFDs discovered from
the HSC imaging data to constrain their luminosities based on the identification of member
stars, as well as their dynamical masses deduced from the velocity distribution of stars.
We have already obtained spectra of several of these candidates. The completion of the HSC-SSP
survey program will provide a more robust constraint on the observed number of satellites,
which will eventually shed light on
the nature of dark matter on small scales, i.e. the scales of galaxies and satellites.

\begin{ack}
This work makes use of data collected at the Subaru Telescope and retrieved from
the HSC data archive system, which is operated by Subaru Telescope and Astronomy
Data Center at National Astronomical Observatory of Japan.
MC thanks support from MEXT Grant-in-Aid for Scientific Research on 
Innovative Areas (No. 15H05889, 16H01086, 18H04334).
NA is supported by the Brain Pool Program, which is funded by the Ministry of Science
and ICT through the National Research Foundation of Korea (2018H1D3A2000902).

The HSC collaboration includes the astronomical communities
of Japan and Taiwan, and Princeton University. The HSC instrumentation and
software were developed by the National Astronomical Observatory of Japan (NAOJ),
the Kavli Institute for the Physics and Mathematics of the Universe (Kavli IPMU),
the University of Tokyo, the High Energy Accelerator Research Organization (KEK),
the Academia Sinica Institute for Astronomy and Astrophysics in Taiwan (ASIAA),
and Princeton University. Funding was contributed by the FIRST program from Japanese
Cabinet Office, the Ministry of Education, Culture, Sports, Science and Technology,
the Japan Society for the Promotion of Science, Japan Science and
Technology Agency, the Toray Science Foundation, NAOJ, Kavli IPMU, KEK, ASIAA,
and Princeton University.  

This paper utilizes software developed for the LSST.
We thank the LSST Project for making their code freely available
at \url{http://dm.lsst.org}.

The PS1 Surveys have been made possible through contributions of
the Institute for Astronomy, the University of Hawaii, the Pan-STARRS Project Office,
the Max-Planck Society and its participating institutes, the Max Planck Institute for Astronomy,
Heidelberg and the Max Planck Institute for Extraterrestrial Physics, Garching,
The Johns Hopkins University, Durham University, the University of Edinburgh,
Queen's University Belfast, the Harvard-Smithsonian Center for Astrophysics,
the Las Cumbres Observatory Global Telescope Network Incorporated,
the National Central University of Taiwan, the Space Telescope Science Institute,
the National Aeronautics and Space Administration under Grant No. NNX08AR22G issued through
the Planetary Science Division of the NASA Science Mission Directorate,
the National Science Foundation under Grant No. AST-1238877, the University of Maryland,
and Eotvos Lorand University and the Los Alamos National Laboratory.
\end{ack}


\end{document}